# Enhancing Psychometric Analysis with Interactive SIA Modules

A Preprint

**Patrícia Martinková**[1,2], **Jan Netík**[1,2], **Adéla Hladká**[1]
[1] Institute of Computer Science of the Czech Academy of Sciences, Prague, Czech Republic
[2] Faculty of Education, Charles University, Prague, Czech Republic

This is the authors' version of the manuscript.

October 31, 2025

## Abstract

`ShinyItemAnalysis` (SIA) is an R package and `shiny` application for an interactive presentation of psychometric methods and analysis of multi-item measurements in psychology, education, and social sciences in general. In this article, we present a new feature introduced in the recent version of the package, called "SIA modules", which allows researchers and practitioners to offer new analytical methods for broader use via add-on extensions. SIA modules are designed to integrate with and build upon the SIA interactive application, enabling them to leverage the existing infrastructure for tasks such as data uploading and processing. They can access and further use a range of outputs from various analyses, including models and datasets. Because SIA modules come in R packages (or extend the existing ones), they may come bundled with their datasets, use object-oriented systems, or even compiled code. We illustrate the concepts using sample modules from the newly introduced `SIAmodules` package and other packages. After providing a general overview of building Shiny applications, we describe how to develop the SIA add-on modules with the support of the new `SIAtools` package. Finally, we discuss possibilities of future development and emphasize the importance of freely available, interactive psychometric software for dissemination of methodological innovations.

## 1 Introduction

Measurement in social sciences diverges from the straightforward quantification of physical attributes such as height or weight because the measured traits are *latent*, existing beyond direct observation. Such measurements consist of a considerable amount of error, which needs to be accounted for, and they often involve multiple raters and/or multi-item instruments. Consequently, a spectrum of statistical and psychometric models and techniques has been developed to analyze measurements in the social sciences (Rao & Sinharay, 2007; Bartholomew, Steele, & Moustaki, 2008; Martinková & Hladká, 2023). These methodologies include providing proofs of the measurement reliability, and assessing validity by analyzing relationships with criterion variables and by analyzing the internal structure with factor analysis. Given that social science measurement typically entails multiple components such as items (or criteria, raters, occasions, etc.), there is a particular focus on modeling item responses within multi-item measurements and checking the functioning of each individual item.

To make these complex analytical methods more accessible, psychometric researchers often implement newly proposed models and algorithms in freely available programming software such as R, including its widely used packages such as `mirt`(Chalmers, 2012), `lavaan`(Rosseel, 2012), or `psych`(Revelle, 2024). This may, however, still limit the pool of users to those with programming skills.

To further expand the range of the new methods to a broad audience, software tools have been developed that integrate statistical computing with interactive interfaces. One such enabling technology is Shiny, an open-source framework for building web applications in R or Python (Chang et al., 2024; Wickham, 2021). It enables users to create dynamic user interfaces and responsive visualizations directly from their statistical code, without requiring prior knowledge of web technologies such as HTML, CSS, or JavaScript. This feature makes Shiny particularly appealing to data analysts and researchers proficient in R, as it lowers the barrier to developing and deploying web-based tools for data exploration and analysis.

Building on this framework via the **`shiny`** package for building interactive web applications directly from R (Chang et al., 2024), the **`ShinyItemAnalysis`** (SIA) package (Martinková & Hladká, 2023; Martinková & Drabinová, 2018) was developed to provide a collection of psychometrics tools through both a Command-Line Interface (CLI) and an interactive application. The interactive SIA application (Figure 1) includes access to various example datasets, supports uploading and analyzing user data, and enables users to download tables and figures. It also offers automated PDF and HTML report generation, helping to integrate psychometric analysis into the test-development process.

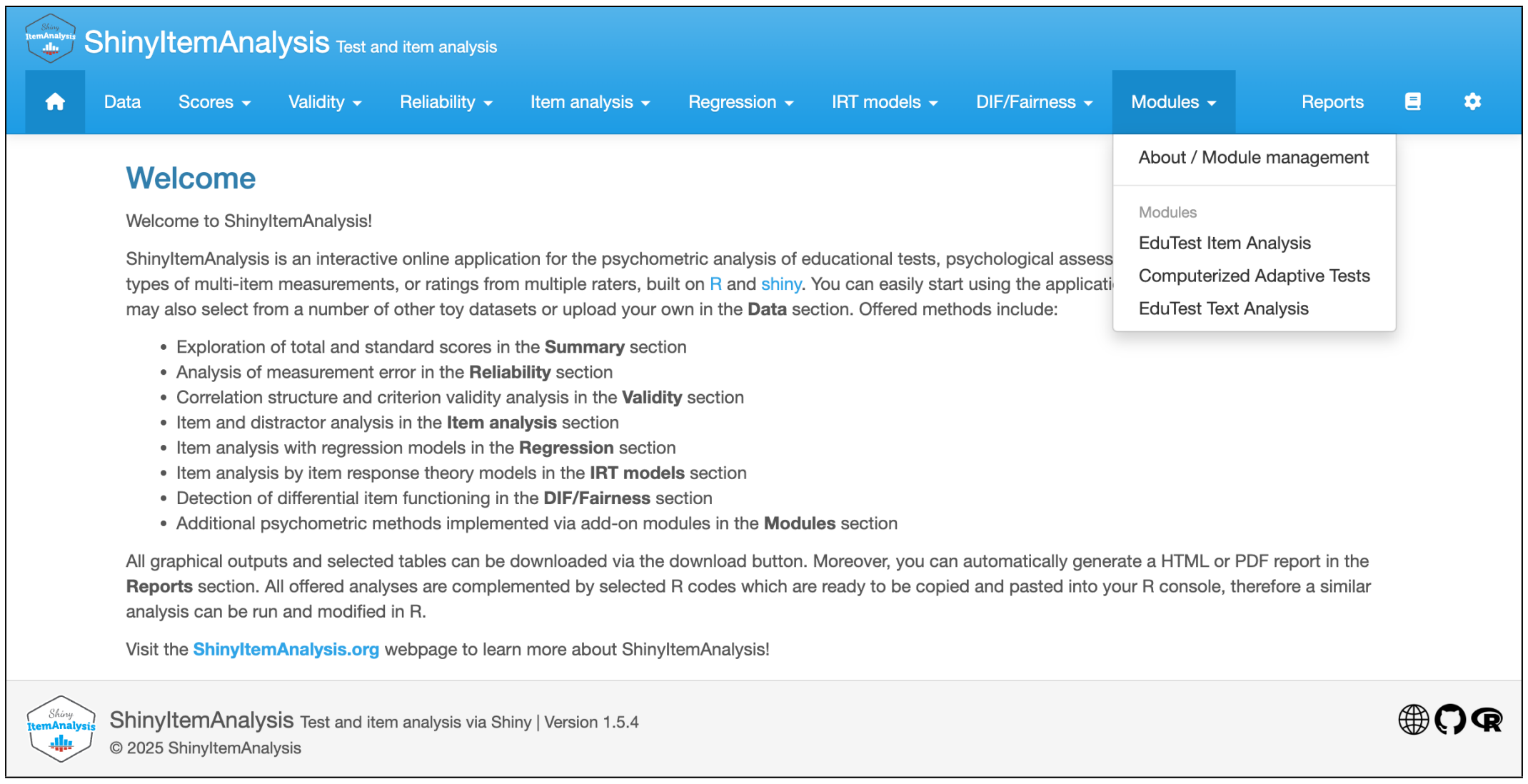

Figure 1: ShinyItemAnalysis application introduction screen

SIA's interactive environment supports both teaching and applied use of psychometric methods, making them more accessible to a broad audience. It includes training sections with automatically graded exercises and serves as an entry point for users new to R. Sample code is provided to bridge the gap between the Graphical User Interface (GUI) and the CLI, demonstrating how to apply SIA functions alongside standard psychometric packages such as **`mirt`** (Chalmers, 2012), **`psych`** (Revelle, 2024), and **`difNLR`** (Hladká & Martinková, 2020). Moreover, the application runs as a background job, allowing for using the R console for practicing or modifying selected code at the same time as the application.

The psychometric analyses within the application are structured in sections, aligning with the workflow outlined in the *Standards* (AERA, APA, and NCME, 2014) and more closely described by Martinková and Hladká (2023). The main SIA application includes fundamental psychometric models and methods for evaluating multi-item measurement in social sciences, covering evidence of measurement validity, models to assess reliability, and diagnostics for individual item functioning. Further, the application incorporates Clasical Test Theory (CTT) and traditional item analysis as foundational methods, and extends these by offering regression models to describe item characteristics and to support the step-by-step development of Item Response Theory (IRT) modeling (Figure 2). Additionally, SIA provides a comprehensive set of tools for Differential Item Functioning (DIF) analysis, and also covers other topics, such as Computerized Adaptive Testing (CAT) or text analysis via add-on modules. Together, these features make SIA a powerful platform for complex psychometric analysis.

Since its release, SIA has seen growing adoption among psychometric researchers, educators, and test developers. Its user-friendly, open-source platform has made it a widely used tool for teaching psychometric concepts and promoting reproducible analyses. While the current user base is concentrated in academia, SIA has also been adopted by test developers, including a national testing agency, to support the development of school admission and leaving exams. The online version of the application has been accessed over 60,000 times from more than 100 countries, and the

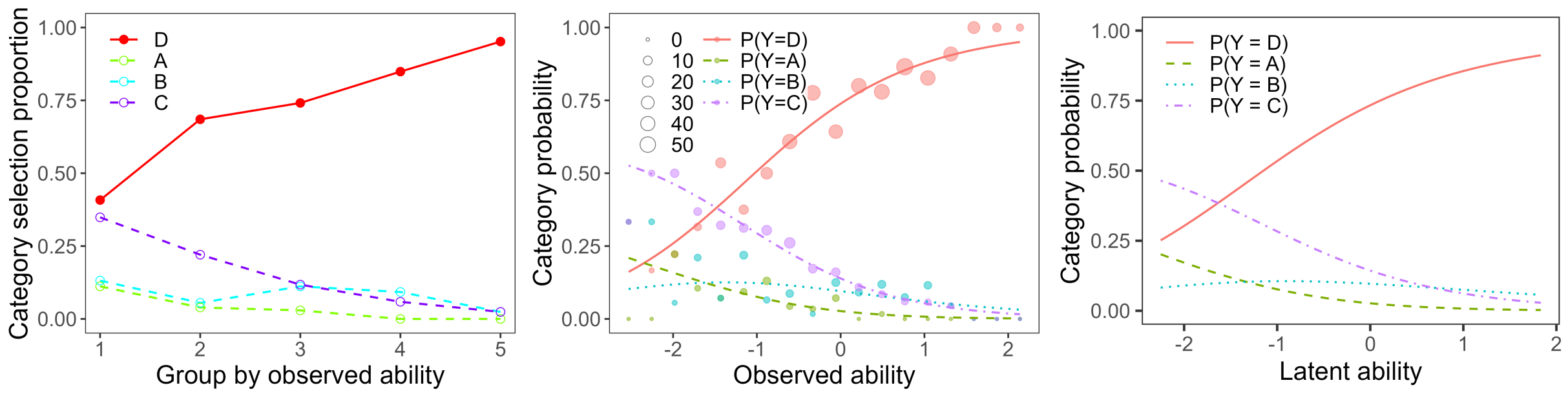

a. Traditional item analysis b. Regression model c. IRT model

Figure 2: Different approaches to item analysis on the same item

package has been downloaded more than 100,000 times from Comprehensive R Archive Network (CRAN), with an additional approximately 1,000 monthly downloads from GitHub. Thanks to its ongoing development and newly introduced modular design, SIA holds strong potential for broader applications in psychometric consulting or health-care-oriented measurements.

Compared to existing psychometric software, SIA offers a well-integrated and modern interface for both classical and modern test analysis. While some methods, such as Structural Equation Modeling (SEM), are not yet implemented in SIA, the application is distinguished by its open-source nature, active and ongoing development, and high extensibility at the source-code level (see Table 1 for comparison).

| Software | Open source | R integration | Extensibility | Implemented methods | | | |
|---|---|---|---|---|---|---|---|
| | | | | CTT | IRT models | SEM | DIF |
| `ShinyItemAnalysis` | Yes | Yes | Full code-level customization, add-ons via `SIAtools` | Yes | Dichotomous, polytomous | No | Yes |
| `jMetrik` | Yes | Yes (`jmetrik`) | Minimal, source code extension only | Yes | Dichotomous, polytomous | No | Yes |
| `WINSTEPS` | No | Partially (`r2Winsteps`) | Limited scripting, input file customization | Yes | Dichotomous (Rasch), polytomous (RSM, PCM) | No | Yes |
| `IRTPRO` | No | Partially (`irt`) | Basic scripting for batch processing | No | Dichotomous, polytomous, multidimensional | No | Yes |
| `Mplus` | No | Yes (`MplusAutomation`) | Flexible model syntax, R workflow | No | Dichotomous, polytomous, multidimensional | Yes | Partial |
| `flexMIRT` | No | Yes (`irtQ`) | Scripting and model specification | No | Dichotomous, polytomous, multidimensional | No | Yes |

Table 1: Comparison of interactive psychometric software tools.

As the field and data complexity undergo a rapid evolution, the ability to adapt and expand analytical tools becomes increasingly important. In response to these demands, we introduce the "SIA modules" framework, a system designed to enable researchers and practitioners to develop add-on SIA modules that seamlessly integrate with and expand upon the capabilities of the main application. In doing so, we take inspiration from `jamovi` (The jamovi project, 2024), `JASP` (Love et al., 2019), and R packages `Rcmdr` (Fox, 2005), `Deducer` (Fellows, 2012), and `RKWard` (Rödiger, Friedrichsmeier, Kapat, & Michalke, 2012), all of which offer extension frameworks, thus being similar to our endeavor. Apart from the aforementioned software, both the SIA application and SIA modules are implemented entirely in the R language, keeping it open for the wider R community.

The remainder of this paper is structured as follows: Section 2 explains how SIA modules differ from standalone Shiny applications in their structure and connectivity with the main SIA application. It also offers an overview of existing modules housed in the `SIAmodules` package (Martinková & Netík, 2024), and examples of modules residing in other packages. Section 3 provides a basic overview of building Shiny applications, together with a demonstration of a standalone Shiny application for CAT simulation. In Section 4, we present a comprehensive guide to SIA module development, including step-by-step instructions. We describe the architecture of SIA modules, how they interact with the core application, what the differences are from the standalone Shiny applications, and we detail the process of developing new modules with the aid of the `SIAtools` package (Netík & Martinková, 2025). Finally, Section 5 provides a discussion on the implications of SIA modules and the potential for broader application of the concept across other research areas.

## 2 SIA modules

While the core SIA application already provides a comprehensive environment for psychometric analyses, its functionality is now being extended through a new modular framework. The recently introduced SIA modules allow researchers and developers to build add-on extensions that integrate directly with the main SIA application. Each module can either introduce new analytical methods or enhance existing ones while fully utilizing the application's infrastructure for data handling, model estimation, and visualization. This design enables both flexibility and scalability—modules can rely on outputs from the core SIA analyses (e.g., IRT models, factor analysis, or DIF detection), use their own datasets, or even implement computationally intensive procedures based on compiled code. Together, the modular architecture establishes an open, collaborative platform that encourages community contributions and facilitates the dissemination of advanced psychometric tools in an interactive and reproducible form.

### 2.1 Running SIA with SIA modules

SIA can be run either online[1] or locally within R. To run locally, the **ShinyItemAnalysis** package and its dependencies have to be installed in R. The application can then be launched by entering the following single command in the R console:

```
ShinyItemAnalysis::run_app()
```

The function will first check whether all dependencies of the interactive application are satisfied, and offer their installation if not. In this simulated example, the **difNLR** R package was not installed:

```
The package "difNLR" (>= 1.5.0) is required to run the app.
Would you like to install it?

1: Yes
2: No
```

Once all the dependencies are resolved, the `run_app()` function also checks if there are any SIA modules available on the official SIA repository[2] that are not yet installed and offers to install them. In this case, the function offers a menu of available packages containing SIA modules, among others, the **SIAmodules** package, which retains a collection of individual modules:

```
Additional SIA modules are available on the official SIA repository
within the following packages.
Would you like to install any of them?
This message is displayed once per session.
See '?ShinyItemAnalysis_options' for more control.

1: All
2: None
3: EduTestItemAnalysis
4: EduTestTextAnalysis
5: IRR2FPR
6: SIAmodules
```

This check is only carried only once per R session and can be disabled completely by setting `options(sia.offer_modules = FALSE)`. Moreover, when run locally, the user may install the modules later on in the GUI as shown in Figure 3. They will become immediately usable without any need to restart the application. If the application runs in the background in a separate R process and a package that contains a SIA module is installed in R library, the user can as well use the "Rediscover modules" button located in the Settings section (under the cogwheel icon in the top right corner) to make the newly installed module available without closing the application.

[1]Recent main mirror is https://shiny.cs.cas.cz/ShinyItemAnalysis, see project webpage https://www.shinyitemanalysis.org/.
[2]https://shinyitemanalysis.org/repo

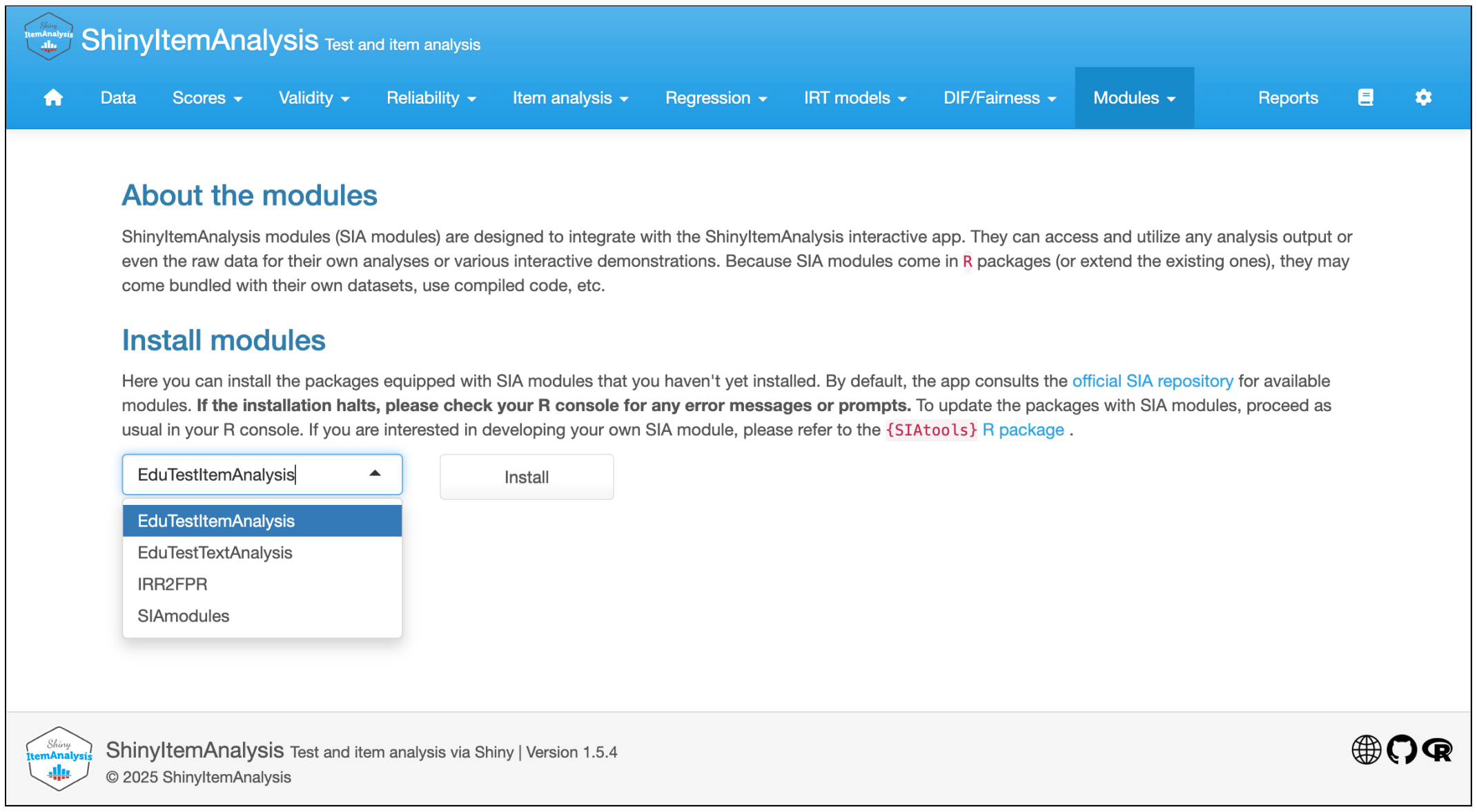

Figure 3: Installing SIA modules in the interactive application

## 2.2 Sample SIA modules

The sample SIA modules presented in this section illustrate the range of possible extensions of the core application. Some operate solely on their own datasets, while others allow interaction with data from the main SIA application or even generate new datasets to be passed back into it.

### 2.2.1 EduTest Item Analysis: Tailoring data upload, adding data and model complexity

The EduTest Item Analysis module within the eponymous `EduTestItemAnalysis` package (Netík & Martinková, 2024) is specifically tailored to accommodate the format of publicly available Czech Matura Exam data (see the supplementary material for sample item data and metadata). It serves as an example of how to create a customized data upload procedure that diverges from the main SIA application. Notably, the EduTest Item Analysis module enables users to upload test metadata specifying the type of each item, i.e., some items may follow the three-parameter logistic IRT model (Birnbaum, 1968) or its restricted Two-Parameter Logistic (2PL) version, some may be modeled by the generalized partial credit model (Muraki, 1992), and some by the nominal response model (Bock, 1972), wherein the correct response must also be specified. This demonstrates the capability to extend the IRT analysis within the SIA application, which currently supports only tests with a single item type. The item-specific IRT modeling is provided in the respective tab of the module, alongside some customized traditional item analyses. Additionally, the module offers the functionality to create binary grouping and/or criterion variables from a factor variable with multiple levels, which can be utilized for DIF detection within the SIA application (Figure 4).

The EduTest Item Analysis module also serves as a non-trivial case study showcasing the module-to-application communication. Upon uploading a dataset to the module, users can modify it to align with the SIA application and reuse it in other tabs beyond the module's scope. In order to be able to run all the analyses of the SIA application and its other additional modules, users are offered to pass data uploaded and edited in the EduTest Item Analysis module directly to the main application via the "Pass data to SIA" button (see Figure 4).

### 2.2.2 CAT module: Utilizing models across the application and its modules

The CAT module (Figure 5) from the `SIAmodules` package simulates an adaptive test with user-defined settings. To start, it generates item responses based on a specified IRT model (dropdown menu) for a respondent of a specified ability (slider). The generated response pattern is presented.

In each step of the CAT post-hoc simulation, the item with the highest information (displayed in the left plot) at the current ability estimate (shown in the right plot) is presented. Based on the respondent's answer (correct/incorrect), the ability estimate is updated, and the process repeats until a stopping criterion, defined by the standard error of the ability estimate (left slider), is met.

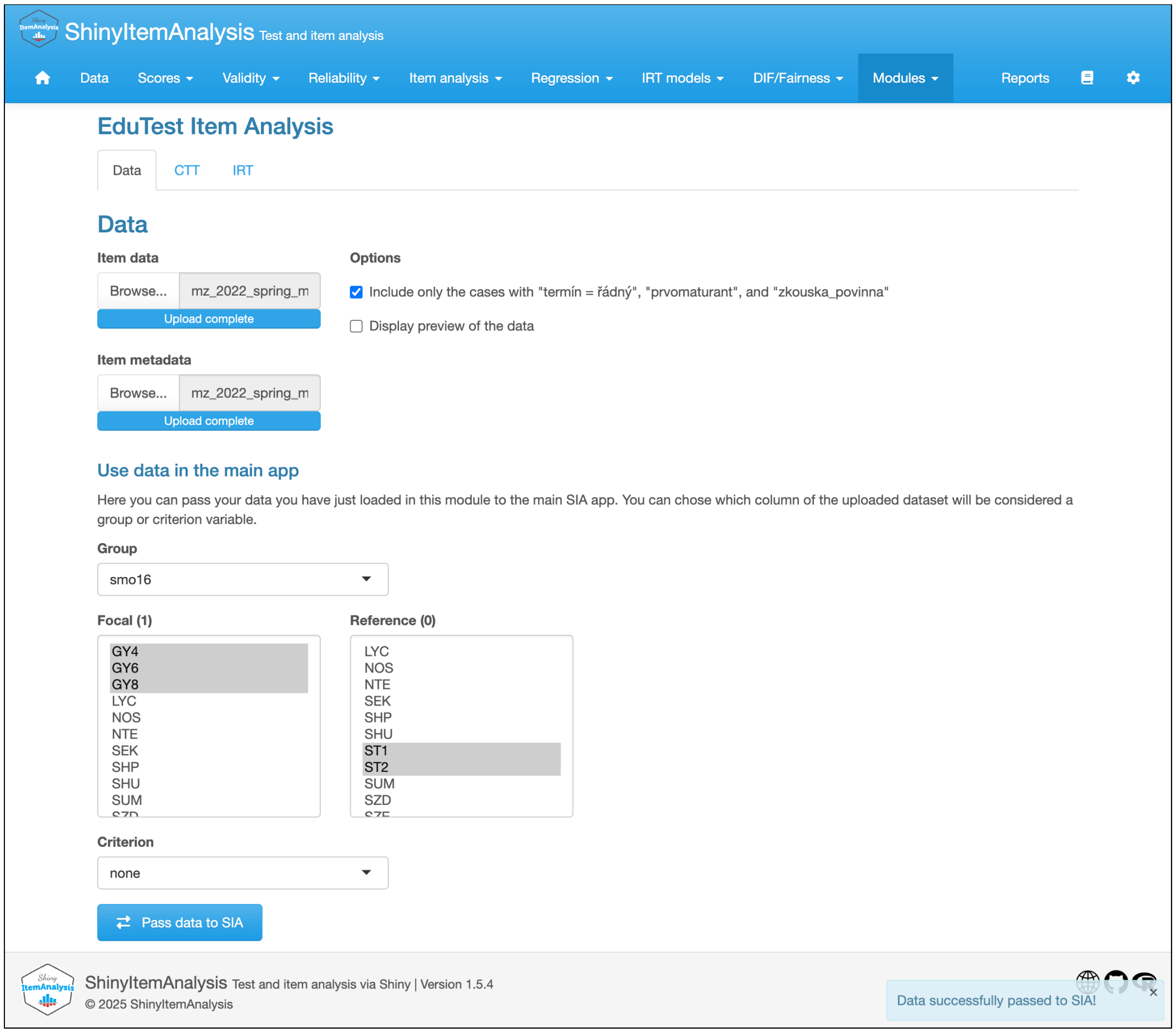

Figure 4: Custom dataset editing in the EduTest Item Analysis module to align with the SIA application format

The module by default offers a 2PL IRT model with predefined item parameters. Moreover, the module offers the possibility to use an IRT model fitted within the main SIA application[3]. This functionality enables interactive CAT simulations on user-uploaded data, whether provided through the "Data" tab or via another module, such as EduTest Item Analysis. This CAT module is a fully developed version of the simplified example used for the in-depth demonstrations in Sections 3 and 4.

### 2.2.3 DIF-C: Extending the DIF analysis to a longitudinal setting

Regression-based methods for DIF detection provide the flexibility to incorporate external matching criteria, such as pre-test scores. Consequently, these models can be used to identify so-called DIF in change (DIF-C; Martinková, Hladká, & Potužníková, 2020) and to analyze item-level heterogeneous treatment effects (Gilbert, 2024). For instance, in the study on learning competencies (Martinková et al., 2020), no overall differences in total scores were observed between students from basic and academic school tracks in either the 6th grade or the 9th grade. Nevertheless, when prior knowledge (6th-grade learning competency scores) was taken into account, certain items in the 9th grade still exhibited differential functioning between the tracks (Figure 6), as identified using the logistic regression model for DIF detection (Swaminathan & Rogers, 1990).

While the DIF-C detection is accessible in the main application using the Learning To Learn 9 toy dataset with the score from the 6th grade as an "Observed score" variable, the DIF-C module from the `SIAmodules` package opens up the core analysis of the study (Martinková et al., 2020) in an extended and directly reproducible way. It provides a step-by-step examination of both scores, a summary of the DIF-C analysis, and plots of Item Characteristic Curves (ICCs) for individual items.

[3]Note that there is no need to fit the models in the "IRT models" tab beforehand. As soon as the reactive expression used by our module is invoked, the selected model is fitted with the defaults provided in the User Interface (UI) of the respective tab. When the underlying data or any relevant inputs change, the model automatically refits.

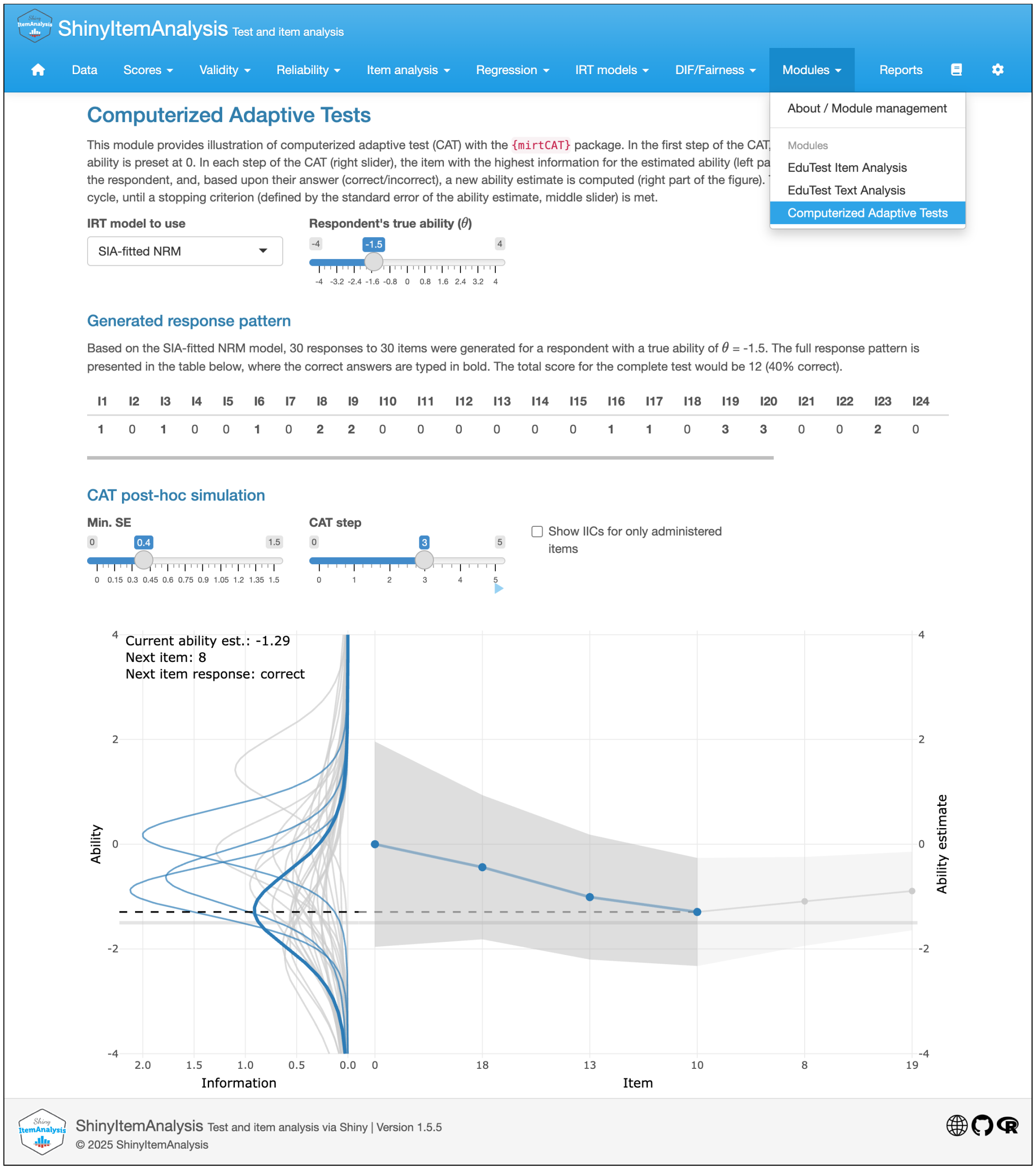

Figure 5: CAT module using the data uploaded in the EduTest Item Analysis module and utilizing the IRT model fitted by the main SIA application

### 2.2.4 Inter-rater reliability: Analyzing ratings from multiple raters

Another aspect of data complexity not currently addressed in the main SIA application involves ratings from multiple raters. When multiple raters are involved, the assessment of Inter-Rater Reliability (IRR) becomes pertinent, typically analyzed through methods such as variance analysis or, more generally, variance component models (Martinková, Bartoš, & Brabec, 2023).

The IRR module (Figure 7) within the `SIAmodules` package provides an interactive demonstration of the issues of using IRR in restricted-range samples in the context of grant proposal peer review. The module demonstrates that when subsets of restricted-quality proposals are used, this will likely result in zero estimates of IRR under many scenarios, although the global IRR may be sufficient (Erosheva, Martinková, & Lee, 2021).

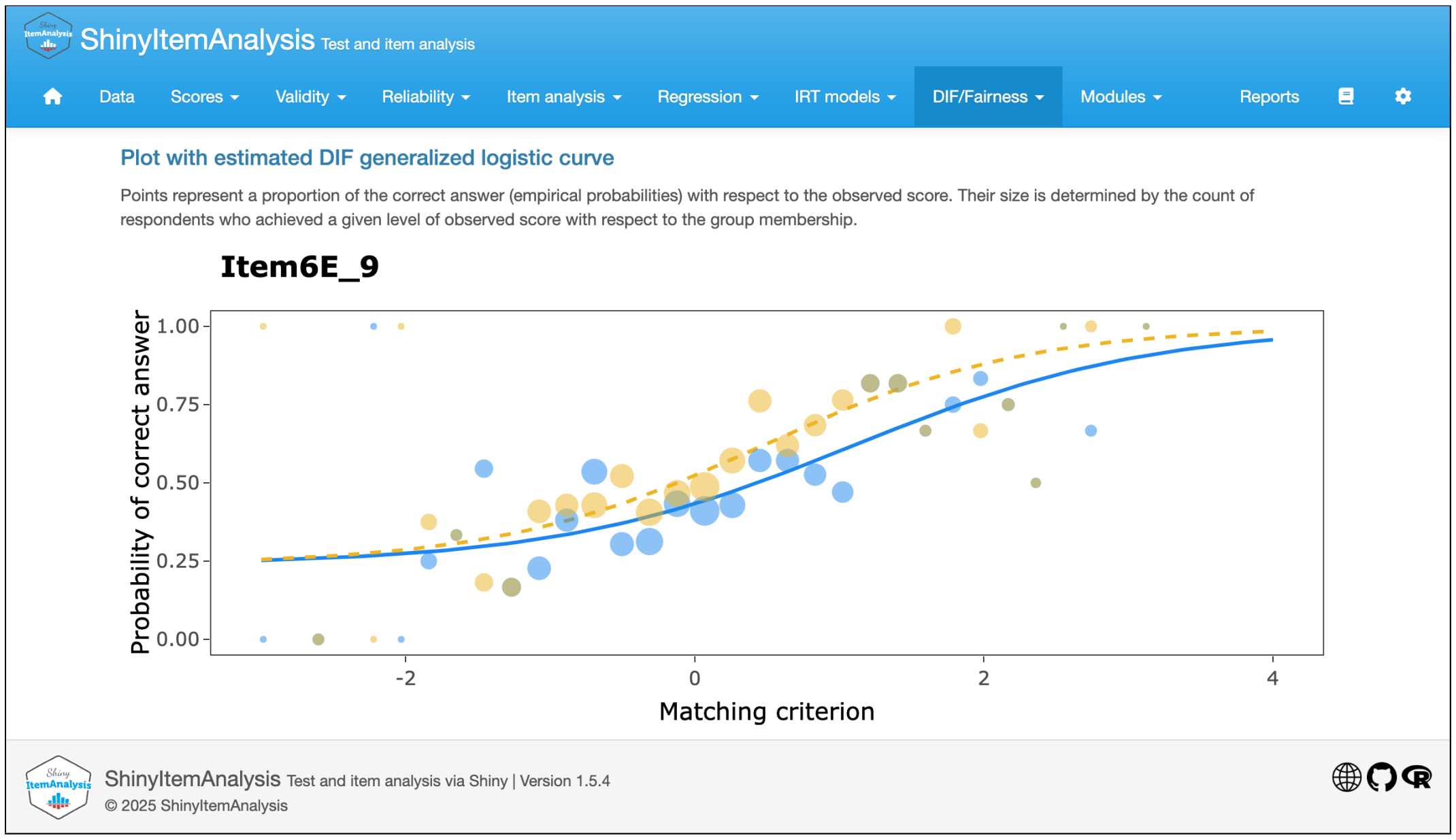

Figure 6: DIF-C module from the `SIAmodules` package

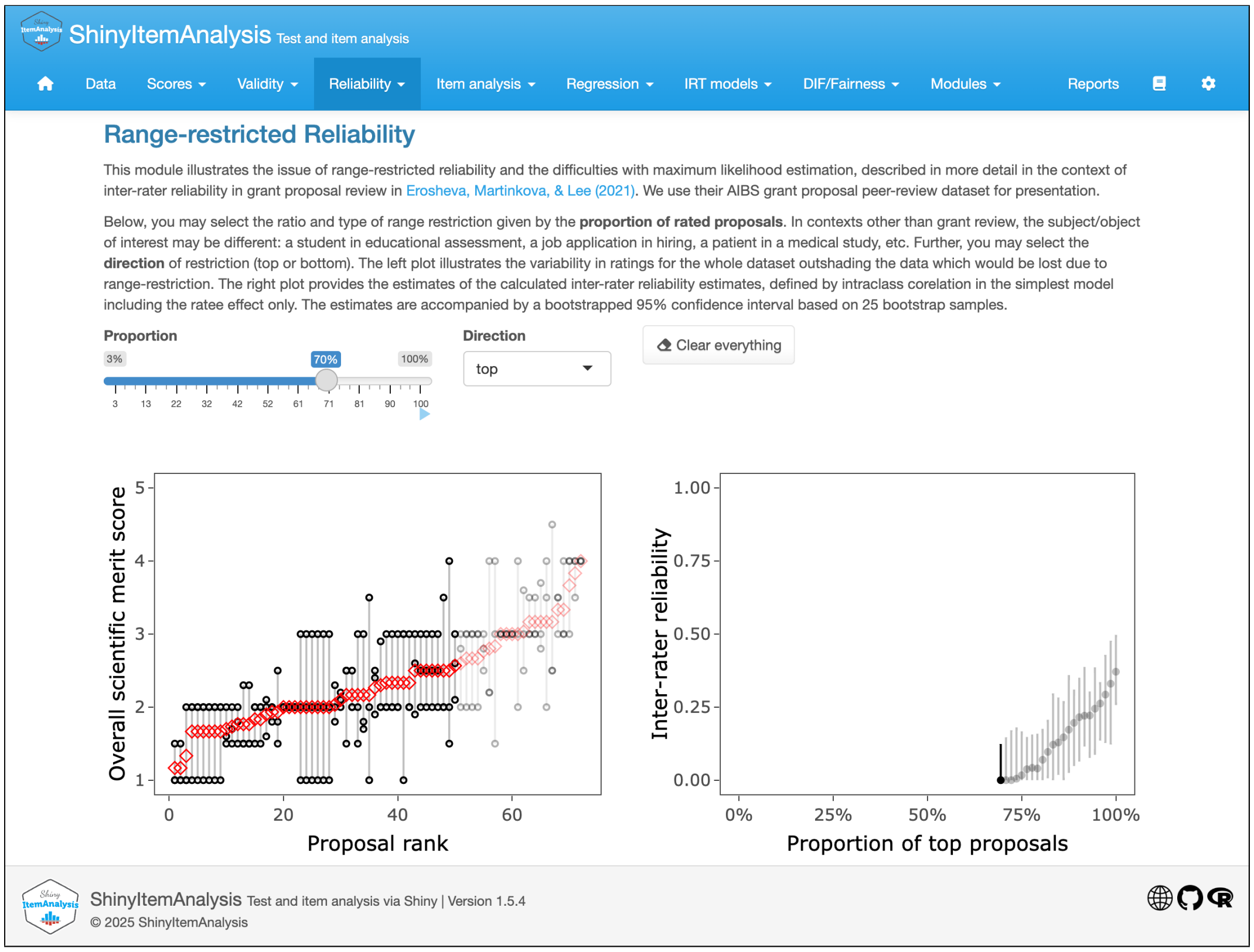

Figure 7: IRR module from the `SIAmodules` package

### 2.2.5 Binary classification metrics from IRR

As another example of a module residing in the “Reliability” tab of the main SIA application, the IRR2FPR module (Figure 8) of the `IRR2FPR` package (Bartoš, 2024) provides an interactive illustration of the calculation of binary classification metrics from IRR, providing an estimate of the probability of correctly selecting the best applicants (Bartoš & Martinková, 2024).

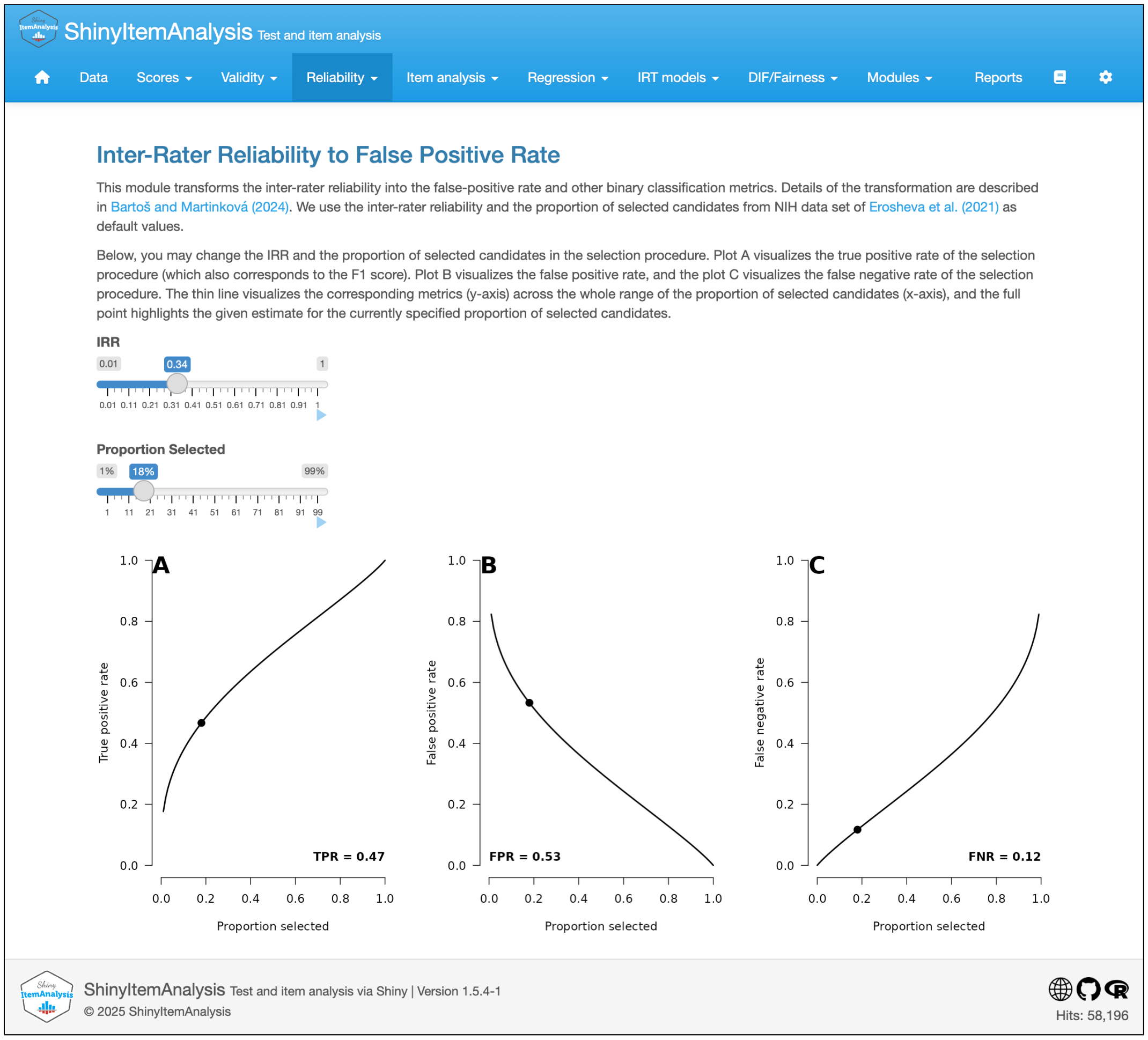

Figure 8: IRR2FPR module from the IRR2FPR package

### 2.2.6 EduTest Text Analysis: Employing large models and compiled code

EduTest Text Analysis module from the `EduTestTextAnalysis` (Figure 9) package (Netík, Dlouhá, Martinková, & Štěpánek, 2024) seeks to provide a tool for item difficulty prediction based solely on the item wording (see Štěpánek, Dlouhá, & Martinková, 2023, for the underlying research). The module does not use any data from the main application, nor does it upload any tabular data. Instead, it uses text input fields and the database of several item examples, demonstrating the versatility of the SIA modules pertaining to the input nature.

Another important feature that this module illustrates is the usage of complex and large models spanning gigabytes of binary data. One of the crucial independent variables in the predictive model is the cosine similarity of different item wording parts (Štěpánek et al., 2023), calculated employing the `word2vec` (Wijffels & Watanabe, 2023) word embeddings model.

In the module, we implemented a mechanism that can download and cache the compressed binary model from the internet on demand and utilize it immediately in the analysis, thus proving that large and complex models are manageable in the proposed modular architecture. The EduTest Text Analysis module also demonstrates the use of compiled C++ libraries that the `word2vec` package is wrapping.

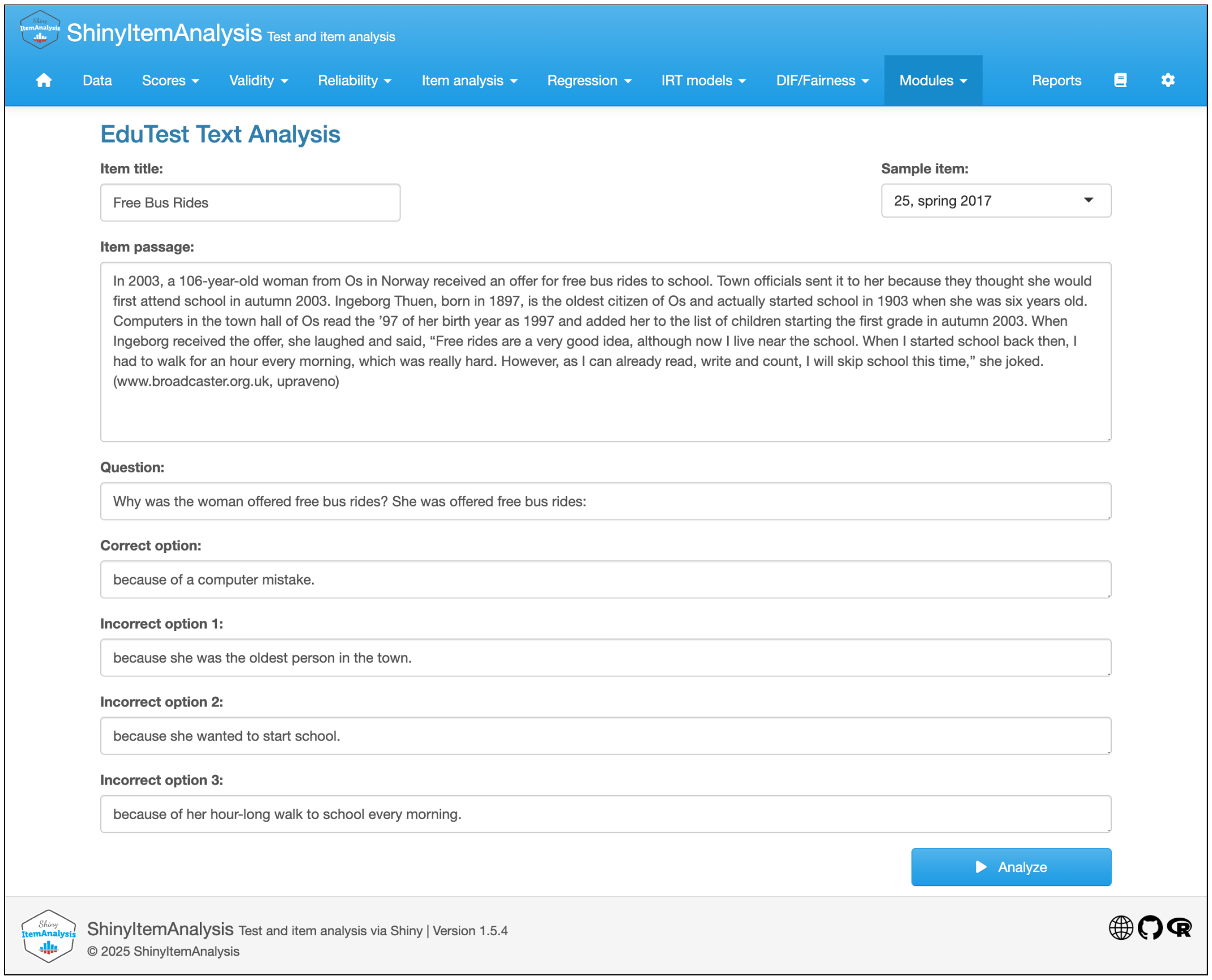

Figure 9: EduTest Text Analysis module from the **EduTestTextAnalysis** package

## 3 Building a standalone Shiny application

Before describing how SIA modules can be developed and integrated with the main application, we provide a general introduction to building standalone Shiny applications. We illustrate the process using the CAT simulation, a simplified example resembling the structure and some functionalities of the CAT module described in Section 2.2.2.

### 3.1 Drafting from R code

Before creating an interactive application, it is often useful to first implement and test the intended functionality as static R code. This ensures that the analysis works as expected before adding the complexity of a Shiny interface. To begin the CAT simulation example, we first load all required libraries.

```
library(mirt)
library(mirtCAT)
library(ShinyItemAnalysis)
```

The CAT simulation relies on an IRT model, for which we consider two options. In the first option, a 2PL IRT model is fitted to the example dataset using the `mirt()` function. Here we consider the HCI dataset (McFarland et al., 2017; Martinková et al., 2017) from the **ShinyItemAnalysis** package, which comprises 20 dichotomously scored items from the Homeostasis Concept Inventory, completed by 651 students (405 males, 246 females).

```
data_2pl_mod <- mirt(data = HCI[, 1:20], itemtype = "2PL")

# display the model parameters
coef(data_2pl_mod, IRTpars = TRUE, simplify = TRUE)
```

```
## $items
##              a      b g u
## Item 1  0.851 -1.146 0 1
## Item 2  0.717 -1.723 0 1
## Item 3  1.574 -1.526 0 1
## Item 4  0.347  1.153 0 1
## ...
```

As the second option, we also consider the 2PL IRT model, now with simulated item parameters using the `generate.mirt_object()` function:

```
# set seed for reproducibility
set.seed(1678934)

# number of generated items set to 50
n_items <- 50L

# simulate 2PL IRT model parameters
pars <- data.frame(
  a1 = rlnorm(n_items, .2, .3),
  d = rnorm(n_items, sd = 1.5)
)

# generate the 2PL IRT model
example_2pl_mod <- generate.mirt_object(pars, itemtype = "2PL")

# display the model parameters
coef(example_2pl_mod, IRTpars = TRUE, simplify = TRUE)
## $items
##              a      b g u
## Item.1  1.500  0.751 0 1
## Item.2  1.345  0.652 0 1
## Item.3  1.258  1.004 0 1
## Item.4  1.829  0.740 0 1
## ...
```

Next, the Shiny inputs are defined, including the underlying IRT model (here stored in the `irt_model` object and later to be selectable via a dropdown menu in the intended Shiny app; see Figure 12) and the respondent's ability (stored in the `theta` object and adjustable via a slider in the future Shiny app; see Figure 12).

```
# Define Shiny app inputs
# select IRT model from pre-defined models
irt_model <- example_2pl_mod

# choose respondent ability theta
theta <- 1
```

Further, the response pattern for a respondent with the selected ability `theta` is generated based on the underlying IRT model `irt_model` using the `generate_pattern()` function from the **mirtCAT** package.

```
(response_pattern <- generate_pattern(irt_model, Theta = theta))

##      [,1] [,2] [,3] [,4] [,5] [,6] [,7] [,8] [,9] [,10] ...
## [1,]    0    1    1    1    1    1    0    1    0     1 ...
```

In this example, the simulated respondent with the ability equal to 1 would answer the first item incorrectly, the next five items correctly, and so on, assuming the items were administered in a linear format.

The generated responses are further used in the CAT post-hoc analysis (see Figure 10). In an initial step, before any items are presented to the respondent, their ability is estimated at 0, and the first item is presented to the respondent, choosing the item with the highest information for this ability level. Based on their answer (correct or incorrect, recorded in advance in the generated response pattern), the ability is recalculated from their response and the IRT model. If the standard error of the estimate is above a given threshold (here 0.35), the termination criterion is not yet met, and the cycle repeats. In each subsequent step, the item with the highest information from the remaining pool for the current ability estimate is presented, the ability estimate is updated based on the respondent's answer, and the termination criterion is checked. This process continues until the termination criterion is met or all items are used, at which point the CAT post-hoc analysis ends.

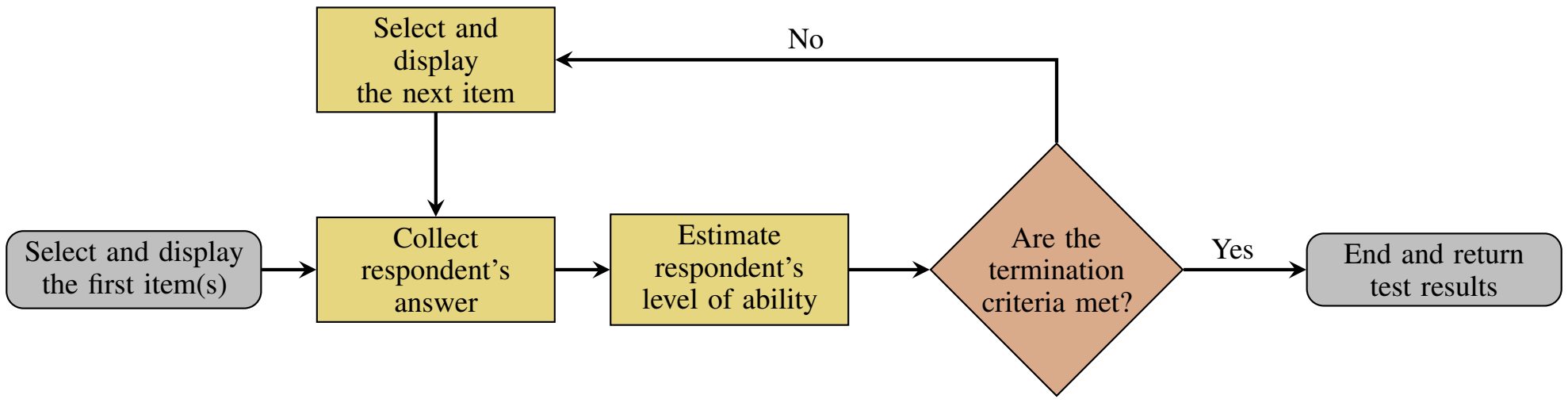

Figure 10: Flowchart of CAT.

The `mirtCAT()` function is used to perform this CAT post-hoc analysis, with the underlying IRT model provided via the `mo` argument, the respondent's response pattern specified in `local_pattern`, the first item selection in `start_item`, subsequent item selection rules set in `criteria`, and additional parameters such as the termination criterion configuration in `design`. Once the post-hoc analysis is completed, the basic results are here printed while the adaptive test trajectory can be visualized using the `plot()` method (see Figure 11).

```
(sim_res <- mirtCAT(
  mo = irt_model,
  local_pattern = response_pattern,
  start_item = "MI", criteria = "MI",
  design = list(min_SEM = .35)
))
##  n.items.answered  Theta_1 SE.Theta_1 True.Theta_1
##                21 1.314722  0.3493085            1

# plot with 95% CI (see ?mirtCAT::plot.mirtCAT for further settings)
plot(sim_res, SE = qnorm(.975))
```

In this case, the CAT post-hoc analysis administered 21 items before meeting the termination criterion. The first item presented was item 12, answered correctly by the simulated respondent, which increased their ability estimate and led to the selection of item 4, also answered correctly. This iterative process continued until the final ability estimate of 1.31, obtained after incorrectly answering item 24, was accompanied by a standard error of 0.349, which was just below the 0.35 threshold, at which point the test concluded.

We now turn to transforming this CAT simulation into an interactive Shiny application. Before we begin with the UI part, we must first load all necessary R packages that will be used in the application. We will work in the R script called `app.R`[4].

app.R

```
library(shiny)
library(mirt)
library(mirtCAT)
```

[4]For the complete application, see the file `app.R` in the Supplementary Materials.

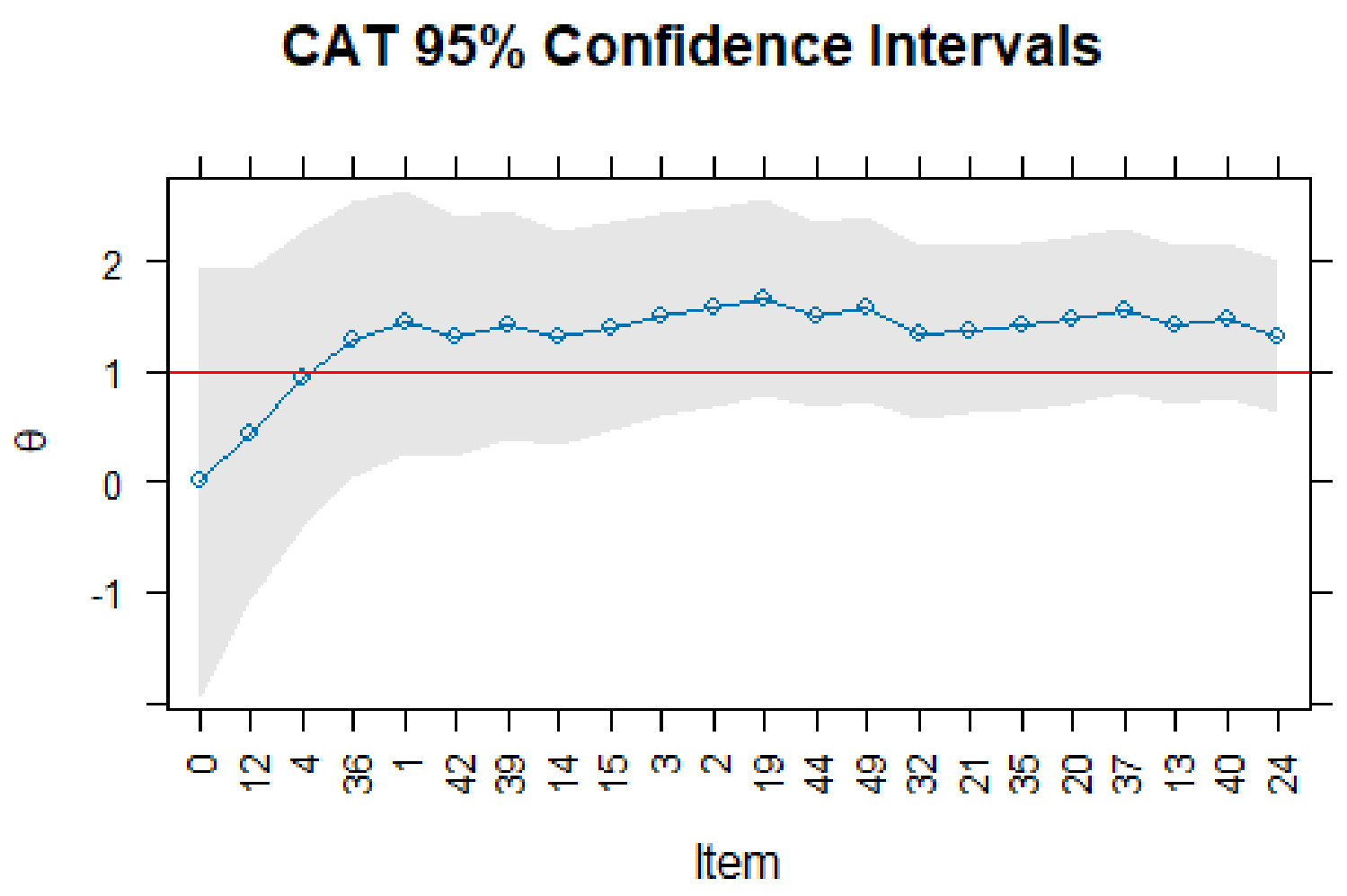

Figure 11: Trajectory of ability estimates with the 95% confidence intervals for a simulated respondent. The red line corresponds to the ability level used to generate the response pattern in the linear setting, which is subsequently used to simulate the test trajectory in an adaptive setting.

```
library(ShinyItemAnalysis)
```

### 3.2 User interface part

The standalone Shiny application consists of two main R objects: the static user interface part and the server function, which are connected through the `shinyApp()` function. The user interface represents the frontend—the part of the application visible to the user in their browser—and is created using the `fluidPage()` layout function, which specifies the overall structure and arrangement of GUI elements.

In our example, the UI begins with a title panel, followed by an introductory paragraph that describes the application's functionality:

app.R

```
ui <- fluidPage(
  titlePanel("Simulation of Computerized Adaptive Test"),
  p("This interactive module simulates a computerized adaptive test (CAT)..."),
```

Next, the dropdown menu is defined using the `selectInput()` function, allowing the user to choose between two 2PL IRT models. By default, the menu is set to the first model with simulated parameters, while the alternative option corresponds to the model estimated from the HCI dataset:

```
  selectInput(
    inputId = "irt_model",
    label = "IRT model to use",
    choices = c(
      "Example 2PL (simulated)" = "example",
      "2PL fitted on HCI data" = "hci"
    )
  ),
```

We also include a slider for setting the respondent's ability, implemented via the `sliderInput()` function. This slider allows values ranging from $-4$ to $4$ in steps of $0.1$, with a default value of 1. Together with the dropdown menu for selecting the IRT model, these inputs enable users to specify the initial settings for the CAT simulation interactively.

```
  sliderInput(
    inputId = "true_theta",
    label = "Respondent's true ability",
    value = 1, min = -4, max = 4, step = .1
  ),
```

Further, the UI part includes a plot output defined using the `plotOutput()` function, which links the CAT simulation results generated in the server logic to the UI (see Section 3.3), allowing users to visualize the adaptive test trajectories directly in the application:

```
  plotOutput("plot"),
```

Finally, a section with references is included at the bottom of the page as an unordered list, and the UI part is closed with a parenthesis:

```
  h3("References"),
  tags$ul(
    tags$li(
      "Martinkov\'a, P., & Hladk\'a, A. (2023). ",
      em("Computational aspects of psychometric methods: With R."),
      " Chapman and Hall/CRC. ",
      a(
        "https://doi.org/10.1201/9781003054313",
        href = "https://doi.org/10.1201/9781003054313"
      )
    )
  )
)
```

In addition to the dropdown menu and slider, a wide variety of interactive widgets, i.e., web elements the user can interact with, can be added into the UI. These include action buttons, checkboxes, radio buttons, numeric inputs, text inputs, and more (Wickham, 2021, Chapter 2). Similarly, beyond the plot output displayed in our example, other reactive outputs, such as text displays or tables, may be included to provide dynamic feedback to users (Wickham, 2021, Chapter 3)[5].

### 3.3 Server part

The server part is the backend of the interactive application – the code that processes user input, runs computations, and returns outputs to be displayed in the browser. It is defined as a function, typically named `server`, with two main required arguments: `input` and `output`. In simple terms, the `input` object is a list that contains all values entered or selected by the user through the UI. The `output` object is a list used to store and return the rendered results, such as plots, tables, or text, back to the UI.

Before we define the server function, we must first create the objects that are not "reactive" – i.e., they do not receive any inputs from the user. In our case, this applies to the IRT models. If we were to define the models within the body of the server function, the code would run unnecessarily every time the application was launched.

```
# seed for reproducibility
set.seed(1678934)
```

[5]See https://shiny.posit.co/r/getstarted/shiny-basics/ for additional options.

```
# number of generated items set to 50
n_items <- 50L

# simulate 2PL IRT model parameters
pars <- data.frame(
  a1 = rlnorm(n_items, .2, .3),
  d = rnorm(n_items, sd = 1.5)
)

# generate the 2PL IRT model based on the simulated parameters
example_2pl_mod <- generate.mirt_object(
  parameters = pars, itemtype = "2PL"
)

# fit the 2PL model on HCI data
data_2pl_mod <- mirt(data = HCI[, 1:20], itemtype = "2PL")
```

Next, we define the server function. Everything that is "reactive" must be specified there. The content of the function can be divided into so-called reactive expressions, which are – in simple terms – functions that automatically update whenever user inputs or any other reactive expressions they depend on change. For instance, the first reactive expression `selected_model()` depends on the model chosen in the UI via the corresponding `irt_model` input. Based on the user's choice, either the 2PL model with generated item parameters (reactive `example_2pl_mod()`) or the model fitted to the HCI dataset (reactive `hci_2pl_mod()`) is retrieved:

```
server <- function(input, output) {
  selected_model <- reactive({
    # return the IRT model based on the selection in the UI
    switch(
      input$irt_model, example = example_2pl_mod, hci = data_2pl_mod
    )
  })
```

Another reactive expression (note that we are still inside the server function body), `sim_res()`, uses both the `selected_model()` and the ability level specified as the `true_theta` input in the UI to generate a response pattern and perform a post-hoc CAT analysis. Finally, the `renderPlot()` function is called on `sim_res()` to visualize the adaptive test trajectory, and the server function is then closed.

```
  sim_res <- reactive({
    # get the selected model
    mod <- selected_model()

    # obtain the response pattern for the given model and selected theta
    response_pattern <- generate_pattern(mod, input$true_theta)

    # perform post-hoc CAT analysis using the selected mode
    mirtCAT(
      mo = mod,
      local_pattern = response_pattern,
      start_item = "MI",
      criteria = "MI",
      design = list(min_SEM = .35)
    )
  })

  output$plot <- renderPlot({
    plot(sim_res(), SE = qnorm(.975))
  })
```

```
}
```

### 3.4 Running Shiny application

To create and run the interactive application, the UI and server parts are connected via the `shinyApp()` function (see Figure 12)

```
shinyApp(ui, server)
```

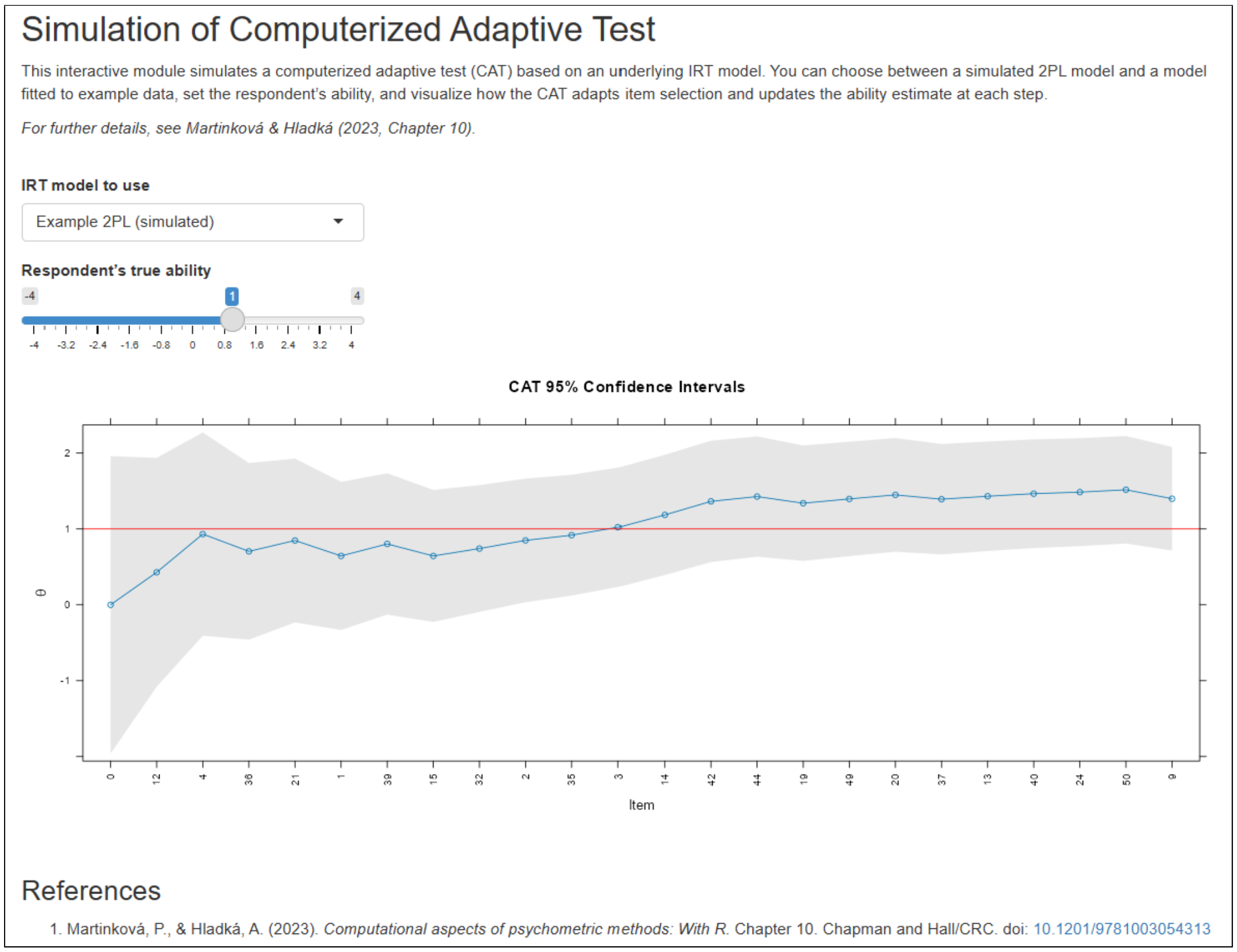

Figure 12: A standalone Shiny application to perform a post-hoc CAT analysis.

## 4 Extending SIA with add-on modules

### 4.1 From a standalone `shiny` application into a SIA module

As outlined above, a standalone **shiny** application is typically developed using either a single `app.R` script or separate `ui.R` and `server.R` files which are then deployed to a Shiny server. Alternatively, for a local, server-free setup, the application can be distributed "as-is" within an R package's `inst` directory. The latter approach differs substantially from the structure recommended by **shiny** authors for larger or long-term projects (Wickham, 2021) and from the approach taken by SIA modules and several other packages related to **shiny** application development, such as **golem** (Fay, Guyader, Rochette, & Girard, 2023) or **rhino** (Żyła et al., 2024).

The main differences from the standalone application approach are as follows: for SIA modules, UI and server parts are both wrapped in dedicated functions that are defined in an R source file inside the `R` directory of the package. The `DESCRIPTION` file of a package declares that it contains one or more SIA modules, and finally, a special YAML file stored within the `inst` directory of the package describes all available SIA modules, including the names of their UI and server functions.

The SIA modules are distributed in ordinary R packages. Developers can create a new package with one or more modules or "append" the modules to their existing packages. Modules available for a package are described in a YAML file, which stores some metadata (title and category) and, crucially, provides bindings to the aforementioned module's functions.

### 4.2 Module discovery and usage

Upon launching the application, a mechanism is dispatched to locate SIA modules within R packages installed in the user's library. Initially, the SIA application collects a list of packages claiming they contain SIA modules. This is declared in the `DESCRIPTION` file of each module package with

```
Config/ShinyItemAnalysis/module: true
```

The following code is used in the `inst/Modules.R` file of the **ShinyItemAnalysis** package to collect the names of available packages containing SIA modules:

```
desc_field <- "Config/ShinyItemAnalysis/module"

names(
  which(
    utils::installed.packages(fields = desc_field)[, desc_field] == "true"
  )
)
```

All such "SIA module" packages are loaded and attached using the standard `library()` call to ensure that any non-trivial features, such as compiled code or S3/S4 methods, work as usual. A reference to the namespace environment of the currently loaded module package is kept in the `ns` object. Subsequently, the YAML file of each module package is searched for the modules' metadata and function bindings. The server function of each module is then located within the package's namespace environment and invoked using `do.call()` with the module's unique identifier as the first argument and with a list of SIA's `reactive`, `reactiveVal`, and `reactiveValues` as the second argument. This enables the module to reuse any reactive object present in the parent application (see Section 4.3 for more details):

```
do.call(ns[[mod_desc$binding$server]], list(id = mod_id, server_dots))
```

The UI of the module function is called in a similar fashion, but inside **shiny**'s `appendTab()` function that appends a new entry to the correct menu:

```
appendTab(
  inputId = "navbar",
  menuName = mod_desc$category,
  tab = tabPanel(
    title = mod_desc$title, value = mod_id,
    do.call(ns[[mod_desc$binding$ui]], list(id = mod_id, ui_dots))
  )
)
```

In the code snippet above, `ns` refers to the package's namespace environment, while `mod_desc` is a list containing the module's metadata, and `mod_id` is the unique module identifier created by concatenating the package's and the module's names (to prevent name collisions).

The category specified in `mod_desc$category` within the module's metadata is used to place the module tab in the desired menu section of the application. Currently, available categories include "Score", "Validity", "Reliability", "Item analysis", "Regression", "IRT models", and "DIF/Fairness". Any other value assigned to the category attribute will automatically position the module into the "Modules" section of the application.

### 4.3 Module development with SIAtools

In order to streamline the development of new SIA modules, we introduce a companion package called **SIAtools** (Netík & Martinková, 2025). This package comprises a collection of functions for constructing and managing modules, along with ready-to-use templates, guides, and various tests to ensure a smooth integration of the module into the SIA application. It is important to note that developers still retain the option to create SIA modules from scratch, as outlined in previous sections. However, with the assistance of **SIAtools**, developers can focus solely on the content of the module itself.

Developers interested in creating a SIA module may find themselves in two distinct situations: (1) they possess an existing R package they wish to extend with one or more SIA modules, or (2) they aim to develop a standalone SIA module. In the latter scenario, the **SIAtools** package offers the capability to create an R package serving as a "container" for the SIA module. This can be achieved either via the R console or through the GUI wizard provided by RStudio. In both cases, developers can utilize the `add_module()` function to configure the package to be compatible with the SIA application. This function automatically generates an entry in the YAML file and creates a template tailored for a new SIA module. Subsequently, developers are encouraged to fill the template with their code, preview the work in progress with the `preview_module()` function, and iterate until satisfied with the outcome.

In the following example, we demonstrate the process of extending the SIA application with the SIA module using the companion **SIAtools** package. We begin by installing the package from CRAN and then loading it:

```
install.packages("SIAtools")
library("SIAtools")
```

To create a SIA module from scratch, one may call `create_module_project("new_proj")`[6] (where `new_proj` represents the name of the new RStudio project that will be created in the current working directory). Another way is to use the RStudio project wizard (File > New project > New Directory > ShinyItemAnalysis Module Project). Upon project creation, a welcome message containing basic instructions will pop up.

For the illustration, we will build a simplified version of the CAT module from the **SIAmodules** package (the resulting simplified module is depicted in Figure 13; the "reference" module is described and shown in Section 2.2.2). Initially, we call `add_module("cat")` to create a new module within the project. This opens the YAML file (also called "SIA Modules Manifest" in the **SIAtools** package), which contains the module specification. Within this file, we modify the title to "CAT Example" and define the module's category (in this instance, "Modules", see Section 4.2). Following these adjustments, the YAML file should resemble the following:

```
sm_cat:
  title: CAT Example
  category: Modules
  binding:
    ui: sm_cat_ui
    server: sm_cat_server
```

The automatically generated module ID (`sm_cat`), located in the first row, serves as the identifier for the module throughout the application. Additionally, the function bindings, which direct SIA to the module's functions, are pre-filled. These functions are housed in `sm_cat.R`, which is automatically created and opened for the developer in RStudio alongside the edited YAML file. In the `sm_cat.R` source file, function skeletons are provided with comments and recommended structure, along with placeholder texts indicating where to insert the code for both the server logic and UI components.

Developers can furnish user-facing documentation at the beginning of the `sm_cat.R` source file. This documentation will be listed along with other functions exported by the package (see `?SIAmodules::sm_cat` for an example of the documentation for CAT module living in the **SIAmodules** package). For instance, the documentation for the simplified example is:

```
# Module documentation -------------------------------------------------
```

[6] The final package is provided as `SampleModulePackage_0.1.0.tar.gz` in Supplementary Materials.

```
# This is the user-facing documentation.

#' CAT Example
#'
#' This SIA module demonstrates basic principles of module development
#' on a simplified version of Computerized Adaptive Tests module from
#' `SIAmodules` package.
#'
#' @author
#' John Doe
#'
#' @references
#' Doe, J. (3024). Sample module. *Journal of Sample Modules, 1*(1),
#' 1--10.
#'
#' @name sm_cat
#' @family SIAmodules
#'
NULL
```

This documentation offers a brief overview of the module's purpose and functionality, enhancing clarity for users accessing the package.

The second part of the `sm_cat.R` file is dedicated to defining the UI function, where developers declare all the components and layout using functions provided by the **shiny** package.

```
## UI part ----------------------------------------------------------

#' `sm_cat` module (internal documentation)
#'
#' [code removed for brevity]
#'
#' @param id *character*, the ID assigned by ShinyItemAnalysis.
#' **Do not set any default value!**
#' @param imports *list* of reactive objects exported by
#'   ShinyItemAnalysis.
#'   See `vignette("imports", "SIAtools")` for more details on how to use
#'   objects from the ShinyItemAnalysis app.
#' @param ... additional parameters (not used at the moment).
#'
#' @keywords internal
#' @rdname sm_cat_internal
#'
#' @import shiny
#'
sm_cat_ui <- function(id, imports = NULL, ...) {
  ns <- NS(id) # shorthand for NS(id, <inputId>)
  # Any `inputId` and `outputId` of {shiny} UI elements MUST be
  # "wrapped" in a `ns()` call! Use `SIAtools::lint_ns()` to check
  # for possible omissions.

  tagList(
    # - - - - - - - - - - - - - - - - - - - - - - - - - - - - - - - - -

    h3("Computerized Adaptive Tests"),
    selectInput(ns("irt_model"),
      "IRT model to use",
      choices = c(
```

```
        "Module's example 2PL" = "example",
        "SIA-fitted IRT for binary data" = "sia_binary"
      )
    ),
    sliderInput(
      ns("true_theta"),
      "Respondent's true ability",
      value = 1, min = -4, max = 4, step = .1
    ),
    plotOutput(ns("plot"))

    # - - - - - - - - - - - - - - - - - - - - - - - - - - - - - - - - -
  )
}
```

Similar to the code presented in Section 3.2, in the code snippet provided above, inside the `tagList()` function, we included a level 3 heading title, an input element used for the selection of an IRT model, a slider input for adjusting the true ability of a respondent, and the final line refers to the output of this sample module – a plot.

There are some notable differences from the code of the standalone application provided in Section 3.2. An important detail to notice is the usage of the `ns()`[7] function around all **shiny** input and output IDs within the UI function. This function is defined directly in the `sm_cat.R` source file and ensures that UI input and output IDs match with those expected by their server logic counterparts. The **SIAtools** package is shipped with a simple *linter* compatible with the **lintr** package (Hester et al., 2023) that checks for the omission of this "namespacing" practice, which can be challenging to debug (for further details, refer to `?module_namespace_linter`).

In the third part of the `sm_cat.R` file, we define the server logic, which, in our case, is done within the `sm_cat_server()` function.

```
## Server part -------------------------------------------------------

#' @rdname sm_cat_internal
#'
#' @importFrom mirtCAT generate_pattern mirtCAT
#'
sm_cat_server <- function(id, imports = NULL, ...) {
  moduleServer(id, function(input, output, session) {
    ns <- session$ns
    # - - - - - - - - - - - - - - - - - - - - - - - - - - - - - - - - -

    mod <- reactive({
      switch(input$irt_model,
        example = example_2pl_mod,
        sia_binary = imports$IRT_binary_model()
      )
    })

    sim_res <- reactive({
      response_pattern <- generate_pattern(mod(), input$true_theta)
      mirtCAT(
        mo = mod(), local_pattern = response_pattern,
        start_item = "MI", criteria = "MI", design = list(min_SEM = .4)
      )
    })

    output$plot <- renderPlot({
```

[7]The name of the `ns()` function is derived from the term "namespace". However, it does not refer to an R package namespace as mentioned in Section 4.2.

```
      plot(sim_res(), SE = qnorm(.975))
    })

    # - - - - - - - - - - - - - - - - - - - - - - - - - - - - - - - - -
  })
}
```

Similar to the code presented in Section 3.3, in the server logic code, we create a reactive expression `mod()` that depends on the model selected in the corresponding input. As was the case in the standalone application presented in Section 3, one of the options is an IRT model with simulated parameters. The second, and default option, however, differs. By default, this SIA module retrieves an example IRT model, previously defined and stored as internal data of the package (see the complete code in file `data-raw/example_2pl_mod.R` in `SampleModulePackage.zip` in the supplementary materials). This allows us to reference the model within our package. The second reactive expression is the output from the simulation of CAT administration itself, based on the response pattern generated using the true latent ability selected with the slider and the IRT model in use. In each step of the iterative CAT algorithm (Figure 10), the simulation will "present" the item with maximal information (MI) for a "current" estimate of the latent ability, and the test will stop when a minimal standard error of 0.40 is reached. In the final segment of the provided code snippet, we call the `plot()` method of the **mirtCAT** package (Chalmers, 2016) to generate a plot of the post-hoc simulation. It is worth noting that every function from external packages must be imported using, for instance, the `@importFrom` tag and declared in the `DESCRIPTION` file because we are constructing a formal R package (Wickham & Bryan, 2023).

To preview work in progress, simply call the `preview_module()` function. By default, all **roxygen2** (Wickham, Danenberg, Csárdi, & Eugster, 2024) blocks[8] automatically convert into the corresponding `.Rd` R manual files and `NAMESPACE` entries. The entire package is loaded and attached without requiring any installation in order to streamline the development as much as possible. The module's functions are called within a simple **shiny** application skeleton (see `?preview_module` for more details), and when executed, the application launches (see Figure 13).

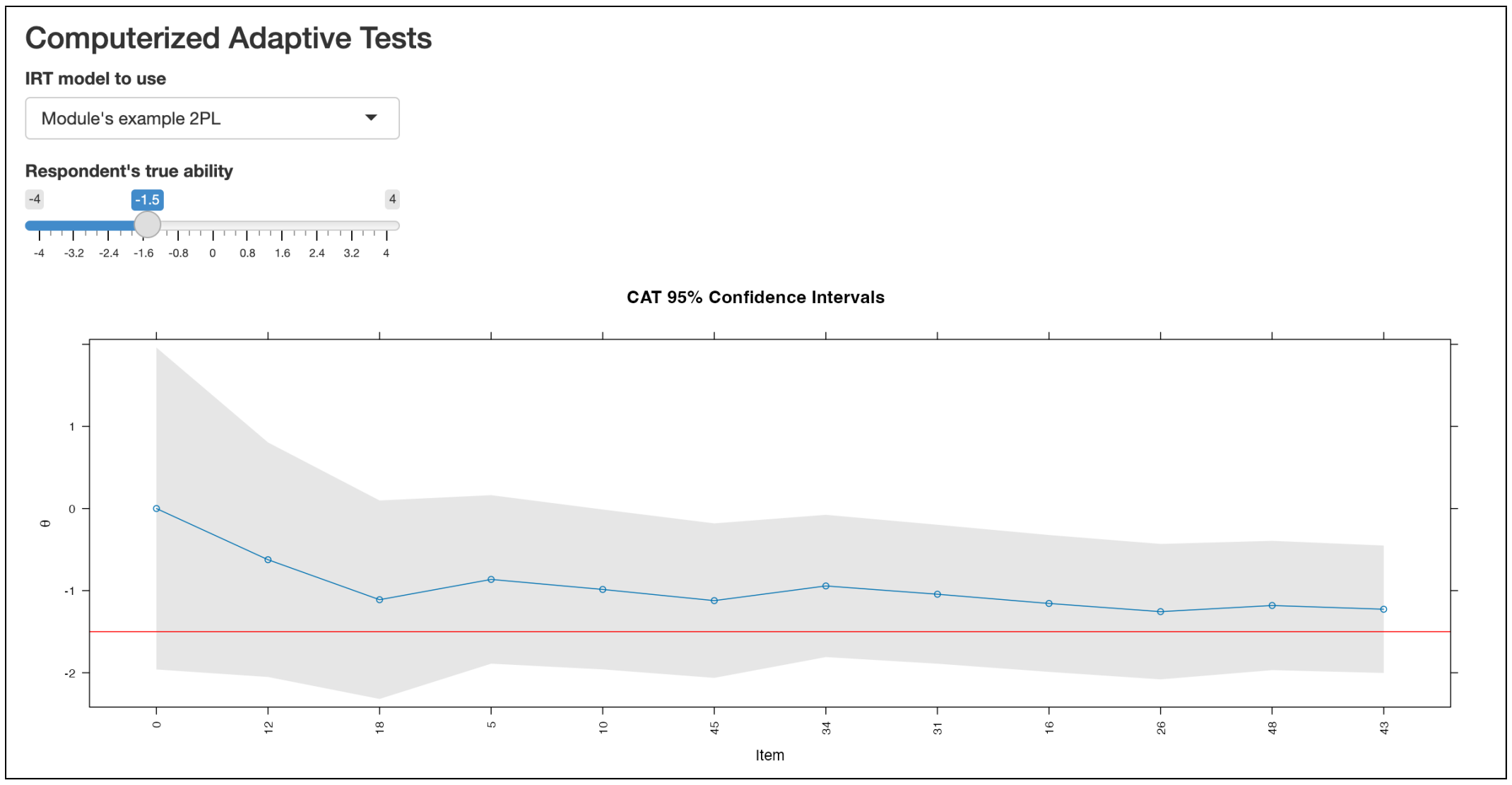

Figure 13: A preview of the sample CAT module

The module, by default, utilizes the aforementioned example IRT model (Figure 13). However, it is important to note that an error message would be displayed if we selected another model from the SIA application. This occurs because—referring to the source code of the module's server logic—the module attempts to utilize the `imports$IRT_binary_model()` reactive, which does not exist. That is because, in the preview, there is no connection to the SIA application, resulting in the `imports` argument of the `sm_cat_server()` function being `NULL`[9].

To test the module fully connected to the parent SIA application, we have to build the package from source and install it as any other R package. Then, once we run the application locally with `ShinyItemAnalysis::run_app()`, our module will appear in the Modules tab (as specified in the YAML file), and we can test it with the IRT model fitted by the SIA application. Now, the module receives the `imports` list of reactives populated by the running SIA application and is able to utilize an IRT model from the SIA application, as was demonstrated in Section 2.2.2.

[8] A sequence of lines starting with `#'`.

[9] Nevertheless, `preview_module()` allows to pass inputs in preview mode, see the documentation.

### 4.4 Module distribution

As outlined earlier, SIA modules are distributed within standard R packages (containing one or more modules). The official repository is located at `https://shinyitemanalysis.org/repo/`. Users can install these packages using the conventional `install.packages()` function and this URL provided in `repos` argument. Since the repository hosts only the packages of interest without any external dependencies, users need to provide their usual CRAN mirror to ensure the proper installation of the module packages. For instance, to install the SIA module contained in the **EduTestItemAnalysis** package, which is not available from CRAN, the following code can be used:

```
install.packages("EduTestItemAnalysis", repos =
  c("https://shinyitemanalysis.org/repo/", getOption("repos"))
)
```

The list of already available module packages with detailed information is available at https://shinyitemanalysis.org/repo/. In the R console, users or developers can obtain the list using the following code:

```
rownames(
  available.packages(repos = "https://shinyitemanalysis.org/repo/")
)
```

## 5 Summary and discussion

In this paper, we have outlined the process of developing new add-on modules for the SIA interactive application in R with a focus on extending tools for psychometric analysis, with the help of the **SIAtools** package. We demonstrated fundamental principles and flexible options within the SIA framework using several modules that are already available and of varying complexity.

Interactive Shiny applications have the potential to expand the user community —particularly applied psychometric researchers and educators— and when R code is provided, these applications lower the barrier to adopting R-based workflows. The modular platform we introduced for add-on packages, utilized in the SIA framework, promotes open collaboration, reproducibility, and customization of **shiny**-based psychometric tools.

The main benefit of the add-on modules lies in their ability to enhance the extensive capabilities and functionalities of the **ShinyItemAnalysis** package, namely with contemporaneous methods and models or didactic showcases. This package offers numerous features, including toy datasets, the option to upload custom datasets, a variety of functions utilized in the interactive application, and the possibility to generate automatic reports, among others. As we have demonstrated with the CAT example module, several modules of the **SIAmodules** package, and with the EduTest Item Analysis and EduTest Text Analysis modules, SIA add-on modules can leverage datasets provided by the SIA application but also use their own ones, up to large complex models or data processed with efficient C++ libraries. Furthermore, they can extend beyond the data types offered in the main SIA application. These modules can seamlessly integrate with the main application; for instance, as we have demonstrated, one module may pass data to the main application for analysis and subsequently transfer the resulting model to another add-on module for further analysis and interactive presentation of the results (e.g., post-hoc analysis of adaptive test), and it does all of this automatically under the hood thanks to reactivity principles of **shiny**.

To the best of our knowledge, only **jamovi** (The jamovi project, 2024) and **JASP** (Love et al., 2019) currently offer the possibility to fully or partially access their analyses programmatically from within the R console. This is enabled by their reliance on underlying R packages. Specifically, the **jmv** package (Selker, Love, & Dropmann, 2023) for **jamovi**, and various packages for **JASP**). Nevertheless, both platforms are primarily designed to be run as a standalone software with GUI in the first place, with programmatic access being a secondary, albeit valuable, feature.

We offer the **SIAtools** package as a resource to facilitate SIA module development. A similar toolkit is provided in the **jmvtools** package (Love, 2024), which also provides a few templates and crucially uses a proprietary compiler to create a module that can be used in the **jamovi** software with a GUI. On top of that, **jmvtools** needs a special

JavaScript runtime environment to operate. This is also the case for another similar package called `jaspTools` (de Jong & van den Bergh, 2024), which relies on a number of external dependencies as well. In contrast, the `SIAtools` works solely within the realm of the R language.

One of the key advantages of the SIA framework is its open-source nature, which aligns with the principles of open science and reproducible research. The `ShinyItemAnalysis` package is licensed under the GNU General Public License version 3 (Free Software Foundation, 2007), a copyleft license that ensures end users' freedom to use, study, share, and modify the software. Under the terms of this license, modules that are derived from or tightly integrated with SIA, such that they form a combined work, are also required to be distributed under GPL-3. Contributors are therefore encouraged to consider licensing compatibility when developing new modules, especially if they rely on tight integration with the core SIA application.

There are several aspects worth considering in the development of the SIA ecosystem and in future versions of the `ShinyItemAnalysis`, `SIAtools`, and `SIAmodules` packages. The `SIAtools` package as well as the broader ecosystem, workflows, and development support for contributors could be further refined in terms of module testing, building, and submitting to maintain rigor and quality in future contributions. While some recent modules, such as EduTest Item Analysis, combine multiple topics to connect specific workflows, e.g., extensions to data upload, traditional item analysis, and more complex IRT models, separate modules positioned in their respective sections of SIA are advisable. In the future, the automatic generation of PDF and HTML reports may offer an editable version that includes outputs from the modules. Further automation of report generation via a command-line environment could foster automation, reproducibility, and reuse with other packages.

While the planned improvements would further enhance the SIA ecosystem, the current suite of `ShinyItemAnalysis`, `SIAtools`, and `SIAmodules` packages already establishes a robust and openly accessible framework. Its graphical user interface can be especially valuable for applied researchers who may lack experience with R or other statistical programming environments, enabling them to conduct advanced psychometric analyses without steep technical barriers. For methodologists and developers, the framework offers a straightforward path to make their methods available interactively, complementing existing R packages with an accessible front end. By fostering openness, not only to users but also to contributors, SIA has evolved into a collaborative ecosystem that can support innovation, reproducibility, and the broader dissemination of psychometric methods.

## Online Supplementary Material

Supplementary material, including the accompanying R scripts, is available at https://osf.io/cnbrh/.

## Acknowledgments

The study was funded by the Czech Science Foundation project 25-16951S "Complex analysis of educational measurement data to understand cognitive demands of assessment tasks", by the project "Research of Excellence on Digital Technologies and Wellbeing CZ.02.01.01/00/22_008/0004583" which is co-financed by the European Union, and by the institutional support RVO 67985807. The authors would like to acknowledge anonymous reviewers for helpful comments on the previous version of the manuscript.